\documentclass[11pt]{article}
\usepackage{epstopdf}                                                          
\usepackage[a4paper,hmarginratio=1:1,vmarginratio=2:3,totalwidth=15.31cm,totalheight=22.55cm]{geometry}
\usepackage{bm,epstopdf,epsfig,amsmath,amssymb,
amsfonts,colordvi,wrapfig,comment,cancel,verbatim,slashed}
\usepackage{graphicx,graphics}
\usepackage[font=md,captionskip=8pt]{subfig}
\usepackage[usenames,dvipsnames]{color}
\usepackage[noadjust]{cite}
\usepackage{xcolor} 
\usepackage[utf8]{inputenc}
\usepackage{setspace}

\usepackage[utf8]{inputenc}

\newcommand{\w}{\omega}  

\allowdisplaybreaks

\newcommand{\cA}{{\cal A}}

\newcommand{\cO}{{\cal O}}

\newcommand{\tr}{\text{tr}}

\newcommand{\be}{\begin{equation}}
\newcommand{\ee}{\end{equation}}
\newcommand{\bea}{\begin{eqnarray}}
\newcommand{\eea}{\end{eqnarray}}
                
\newcommand{\ra}{\rightarrow}  
\newcommand{\Ra}{\Rightarrow}

\newcommand{\baa}{\begin{array}}
\newcommand{\eaa}{\end{array}}

\long\def\symbolfootnote[#1]#2{\begingroup
\def\thefootnote{\fnsymbol{footnote}}\footnote[#1]{#2}\endgroup}

\begin{document} 
\begin{flushright}
\end{flushright}

\thispagestyle{empty}
\vspace{3.cm}
\begin{center}
  \vspace{0.5cm}

 {\Large\bf Unification of Gravity and Standard Model: }

 \bigskip
 {\Large\bf   Weyl-Dirac-Born-Infeld action}

 \vspace{1.5cm}

 {\bf D. M. Ghilencea}
 \symbolfootnote[1]{E-mail: dumitru.ghilencea@cern.ch}
 
\bigskip 

{\small Department of Theoretical Physics, National Institute of Physics
 \smallskip 

 and  Nuclear Engineering (IFIN), Bucharest, 077125 Romania}
\end{center}

\begin{abstract}
  \begin{spacing}{1}
  \noindent
We construct  a unified (quantum) description, by the gauge principle, of
gravity and Standard Model (SM), that generalises  the Dirac-Born-Infeld
action to the SM and Weyl geometry, hereafter  called Weyl-Dirac-Born-Infeld action
(WDBI). The theory is formulated   in  $d\! =\! 4-2\epsilon$ dimensions.
The WDBI action is  a general  gauge theory of SM and Weyl group (of
dilatations and Poincar\'e symmetry), in the Weyl gauge covariant
(metric!) formulation of Weyl geometry. The  theory is SM and Weyl
gauge invariant in $d\!=\! 4-2\epsilon$ dimensions and there is no Weyl anomaly.
The WDBI action  has the unique elegant feature, not present  in other gauge
theories or even in string theory, that it is mathematically
well-defined in $d=4-2\epsilon$ dimensions with no need to introduce in the
action a UV regulator scale or field.
This action actually {\it predicts} that gravity,
through (Weyl covariant) space-time curvature $\hat R$, acts as  UV regulator
of both SM and gravity in $d=4$. A series expansion of the WDBI
action (in dimensionless couplings) recovers in the leading
order a  Weyl gauge invariant version  of SM and the Weyl (gauge theory of)
quadratic gravity. The SM and Einstein-Hilbert gravity  are recovered
in the Stueckelberg broken phase of Weyl gauge symmetry, which
restores Riemannian geometry below Planck scale. Sub-leading orders
are suppressed by powers of (dimensionless) gravitational coupling ($\xi$)
of Weyl quadratic gravity.
\end{spacing}
\end{abstract}

\newpage
\section{Introduction}

In this work we search for a unified (quantum) description,
by the gauge principle \cite{Noether}, of
Standard Model (SM) and gravity. On the SM side this principle was extremely
successful.  Since gravity ``is'' geometry, applying this principle to gravity
means to consider a gauged space-time symmetry; this dictates
the underlying geometry and gravity action as a gauge theory action.
Then what space-time symmetry can we consider beyond Poincar\'e symmetry?

Again, the SM points us in the right direction: SM with
a vanishing Higgs mass {\it parameter} is scale invariant \cite{Bardeen},
which is a hint that this symmetry may be more fundamental\footnote{
Also at high energies or in the early universe,  states
are effectively massless, endorsing this idea.}, in which case we should
gauge it.
This means gauging the Weyl group of dilatations and
Poincar\'e  symmetry \cite{Bregman,Tait,AC,CDA}.
Actually, there is not much else one can do: this is
the  {\it only true gauge  theory} of a space-time symmetry beyond Poincar\'e
\cite{CDA} i.e. with a dynamical/physical gauge boson
\footnote{One can also gauge the larger, full conformal group of
Weyl plus special conformal symmetry \cite{Kaku} to obtain conformal gravity
\cite{Mannheim,Englert}. But this is {\it not} a true gauge theory since its action
cannot have dynamical (physical) gauge bosons of Weyl dilatations and 
of special conformal symmetry \cite{Kaku}, thus this theory does not have the
spectrum of a gauge theory! It is for this reason its action is actually a
{\it particular limit} of the action of the gauge theory of (smaller) Weyl
group discussed here, when the Weyl gauge boson is ``pure gauge''
\cite{Ghilencea:2023wwf,review,AC}.}.
In the absence of matter, the gauge theory of the Weyl group
``is'' Weyl  geometry (WG) \cite{Weyl1,Weyl2,Weyl3} which has
this gauge symmetry by construction: WG is defined
by classes of equivalence of the metric and of Weyl gauge
field ($\omega_\mu$) of dilatations, related by Weyl gauge symmetry transformations.
The associated gravity action is then constructed as a (vector-tensor)
gauge  theory of the Weyl group  \cite{Weyl2}, see \cite{AC,CDA} for an update.

No prior knowledge of Weyl geometry is     needed here.
Given its  gauged dilatation invariance beyond Poincar\'e,
 Weyl geometry can be regarded as Riemannian geometry
 ``covariantised'' with respect to  gauged dilatation symmetry
 (also known as Weyl gauge symmetry) \cite{DG1,review}.
More exactly,  there exists a  Weyl gauge covariant formulation of Weyl
geometry,  which is the only physical formulation and which is
automatically {\it metric} i.e.\,$\hat\nabla_\mu g_{\alpha\beta}=0$ (but non-affine)
\cite{DG1,CDA,CDA2}\footnote{The norm of a vector is invariant
under  Weyl-gauge-covariant parallel transport
\cite{Lasenby,CDA}, \cite{non-metricity} (Appendix~B)},
something  overlooked for a century of
Weyl geometry  despite  Dirac's suggestion \cite{Dirac}.
The associated gauge theory is quadratic in curvatures,
known as Weyl gauge theory of quadratic gravity (``Weyl quadratic gravity''),
see reviews in \cite{review}, \cite{Scholz}.

This  action is spontaneously broken \`a la Stueckelberg \cite{St},
in which the Weyl gauge field of dilatations $\omega_\mu$  acquires mass
proportional to Planck mass $M_p$ after eating the would-be-Goldstone  $\phi$
 (or  ``dilaton'' ghost) propagated by  the higher derivative $\hat R^2$
 term in the action, and  then decouples \cite{Ghilen0,SMW}.
 As a result,  Weyl geometry (connection) becomes Riemannian (Levi-Civita), respectively,
 and  at scales below $M_p$ one recovers
  Einstein-Hilbert action \cite{Ghilen0} with $\Lambda\!>\!0$.
 Thus, the  phase transition where Weyl gauge symmetry is broken
 is interpreted as a change of the underlying geometry.
 No moduli fields are added ad-hoc for this breaking.
 All scales  have geometric origin \cite{SMW,non-metricity}, being related to the vev of 
 field $\phi$ from geometric $\hat R^2$ term, and since this mode was eaten
 by $\omega_\mu$,  its vev does not need to be stabilised.

 With these encouraging results, one can also add matter and consider the
 SM  in  Weyl  geometry, giving the so-called SMW \cite{SMW}. This is an interesting
Weyl gauge invariant theory  that describes gravity and SM,
and respects current  constraints, with
Starobinsky-like inflation \cite{WI1,WI3}, good fits of
galaxy rotation curves \cite{Harko} and black-hole solutions  \cite{Harko2}.
However, this theory (SMW) does not seem the most general one,
since it is ultimately ``gluing'' together in a sum the (Weyl gauge
invariant) actions of SM and of Weyl geometry i.e.\,Weyl quadratic gravity.
It would be good to {\it derive} this action from a more fundamental one.

 The goal of this paper is to find  a more general, unified
 gauge theory action implementing the Weyl gauge symmetry, beyond the SMW scenario.
 The theory should make no distinction between SM and Weyl geometry
 field operators (curvatures, etc), in which these fields {\it and}
 their  derivatives must  transform {\it covariantly} with respect to
 both SM  and Weyl gauge symmetries (much like SM
 fields do with respect to SU(3)$\times$SU(2)$\times$U(1)). External
and internal symmetries must be treated on equal footing in building the action.

Such unified gauge theory can be realized by a version of
Dirac-Born-Infeld action \cite{D,BI,DBI,S,T} due to
{\it both} SM and Weyl geometry, called here Weyl-Dirac-Born-Infeld (WDBI).
The WDBI action is a  space-time integral in $d=4-2\epsilon$ dimensions
of $\sqrt{\det A_{\mu\nu}}$ where $A_{\mu\nu}$
is a linear combination  of operators of mass dimension
2, that are SM and Weyl gauge invariant; these operators are products 
of fields of SM and of Weyl geometry (curvatures) and of their
covariant derivatives. This action gives  a unified framework of internal
(SM) and external (Weyl) gauge symmetries, with manifest covariance/invariance
with respect to both symmetries. This action is obviously more general
than the aforementioned
sum of a Weyl gauge invariant version of SM action and of Weyl geometry
action (Weyl quadratic gravity).

This WDBI action is a truly special gauge theory:  by construction, it is
automatically  SM and Weyl gauge invariant in  $d\!=\! 4-2\epsilon$ dimensions
- a special feature due to Weyl geometry;  the WDBI action is mathematically
well-defined and {\it does not require  an ultraviolet
  (UV) regulator} (be it a DR subtraction scale $\mu$, field, etc)
and all couplings do remain dimensionless.
Introducing a DR scale $\mu$ would  be a big problem since it would actually
 break Weyl gauge symmetry.
The WDBI action  actually {\it predicts} that the Weyl-gauge-covariant\,(!)
space-time curvature $\hat R^\epsilon$  i.e. geometry/gravity acts as 
UV regulator scale/field  in\footnote{
This supports, a-posteriori, the regularisation used in
SMW  \cite{SMW} showing it is Weyl-anomaly free \cite{DG1}.} $d\!=\!4-2\,\epsilon$ 
for SM and gravitational interactions; no DR  scale $\mu$ or field are added by
hand!
 This is  the only gauge theory with this property, showing
 the importance of this  WDBI action.

 This mechanism does not work in  Riemannian geometry where Weyl
 gauge covariance does not exist. For example, in ordinary
 gauge theories a UV regulator scale (DR scale $\mu$, etc) or field
 is required and added by hand. In  conformal gravity  a dilaton  field is
 added by hand as regulator, to maintain its symmetry \cite{Englert}.
Not even in string theory can {\it local} Weyl invariance  (on Riemannian
worldsheet, not in physical space-time as here) be  preserved by regularisation,
being broken by the DR scale $\mu$ that is needed/added in $d=2+\epsilon$;
local Weyl symmetry can then be restored by a condition of vanishing Ricci tensor
in target space \cite{Tong}.

The Weyl gauge invariance  in $d=4-2\epsilon$ dimensions of
the WDBI action of SM and Weyl geometry is important,  since it
implies that this action is automatically Weyl anomaly-free \cite{Duff,Duff2,Duff3,Deser1976,Deser,DG1}.
Hence, the WDBI action is  a consistent (quantum) gauge theory of gravity and SM.
The WDBI action opens a new perspective on physics beyond SM and gravity,
based on the  gauge principle, that goes beyond the
usual {\it quadratic} actions of  gauge theories.

The plan of the paper is this: Section~\ref{WG} reviews
the formalism of WG as a gauge theory. Section~\ref{U} constructs
the WDBI action and shows  how SM and Weyl quadratic gravity
are obtained in a leading order expansion. Einstein-Hilbert action is recovered
in the broken phase, with subleading order corrections suppressed by $M_p$.
 Conclusions are in Section~\ref{C}.

\section{Weyl geometry as a gauge theory of Weyl group}\label{WG}

Let us first review briefly Weyl geometry  as
a gauge theory of the Weyl group, in the Weyl gauge covariant (metric, non-affine)
formulation \cite{DG1}, also \cite{CDA,AC,CDA2} for more details.
The formalism  in this section is actually more  general and
valid in arbitrary  $d$ dimensions, but in the remaining Section~\ref{U}, where the
SM operators and action are added, we obviously have  $d=4-2\epsilon$, ($\epsilon\ra 0$).
Weyl geometry is defined by classes of equivalence of the metric ($g_{\mu\nu}$) and
 gauge field of dilatations ($\omega_\mu$), related by a Weyl gauge transformation,
shown below (in the absence of matter)
\bea\label{WGS}
&\quad&
 g_{\mu\nu}^\prime=\Sigma^2 \,g_{\mu\nu},\quad
 \w_\mu'=\w_\mu -  \partial_\mu\ln\Sigma, 
\quad
\sqrt{g'}=\Sigma^{d} \sqrt{g},
\quad
g^{\prime \mu\nu}=\Sigma^{-2} \, g^{\mu\nu}
\eea

\medskip\noindent
where $\Sigma=\Sigma(x)>0$.
The  Weyl charge $q$ of $g_{\mu\nu}$ was set to $q=2$ - such
normalization for an Abelian symmetry is a choice.
To work with an arbitrary charge for $g_{\mu\nu}$,
and also restore the Weyl gauge coupling $\alpha$, replace 
$\Sigma^2\!\ra\!\Sigma^{q}$ and  $\omega_\mu\!\ra\! (\alpha\,q/2)\,\omega_\mu$
in our results.

Transformation (\ref{WGS}) defines  Weyl gauge symmetry. 
The definition of the geometry is completed by the so-called ``non-metricity''
condition:
\bea\label{GG}
\tilde\nabla_\mu g_{\alpha\beta}+ 2 \, \omega_\mu g_{\alpha\beta}=0,
\quad\textrm{where}\quad
\tilde \nabla_\lambda g_{\mu\nu}\equiv\partial_\lambda g_{\mu\nu} -
\tilde\Gamma^\rho_{\lambda\mu} g_{\rho\nu}-\tilde\Gamma^\rho_{\lambda\nu}g_{\rho\nu}.
\eea

\medskip\noindent
Assuming  a symmetric connection
$\tilde\Gamma_{\mu\nu}^\rho=\tilde\Gamma_{\nu\mu}^\rho$, from (\ref{GG}) one finds
$\tilde\Gamma$, invariant under (\ref{WGS}):
\medskip
\bea\label{tgamma}
\tilde\Gamma_{\mu\nu}^\rho=\Gamma_{\mu\nu}^\rho\big\vert_{\partial_\mu\ra\partial_\mu +
  2\,\omega_\mu}
= \Gamma_{\mu\nu}^\rho+ \big(\delta_\mu^\rho\,\omega_\nu
+\delta_\nu^\rho\omega_\mu -g_{\mu\nu}\omega^\rho\big),
\eea

\medskip\noindent
with $\Gamma$ the familiar Levi-Civita (LC) connection
$\Gamma_{\mu\nu}^\rho=(1/2)\,g^{\rho\lambda} (\partial_\mu g_{\nu\lambda}+\partial_\nu g_{\mu\lambda}
-\partial_\lambda g_{\mu\nu}).$

Further, one associates a Riemann tensor of Weyl geometry $\tilde R^\rho_{\,\,\mu\nu\sigma}$
to $\tilde\Gamma$, via the commutator of two  $\tilde\nabla_\mu(\tilde\Gamma)$ acting
on a vector $v^\rho$:
$[\tilde\nabla_\mu,\tilde\nabla_\nu] v^\rho=\tilde R^\rho_{\,\,\sigma\mu\nu} v^\sigma$.
This gives a Riemann tensor of Weyl geometry, defined by  $\tilde\Gamma$,
by a formula similar to that in Riemannian geometry, but
with $\Gamma$ replaced by $\tilde\Gamma$ \footnote{One has
$\tilde R^\rho_{\,\,\,\sigma\mu\nu}=\partial_\mu \tilde\Gamma^\rho_{\nu\sigma}
-\partial_\nu\tilde\Gamma^\rho_{\mu\sigma}+
\tilde \Gamma^\rho_{\mu\lambda} \tilde\Gamma^\lambda_{\nu\sigma}
-\tilde\Gamma^\rho_{\nu\lambda} \tilde\Gamma^\lambda_{\mu\sigma}$.}.
One then computes the Ricci tensor of Weyl geometry
$\tilde R_{\mu\nu}=\tilde R^\sigma_{\,\,\mu\sigma\nu}$, etc.
This gives the well-known affine, non-metric ($\tilde\nabla_\mu g_{\alpha\beta}\!\not=\!0$)
formulation of Weyl geometry.
This formulation,  used for a century, is not physical since it is {\it not} Weyl
gauge covariant. Indeed, with $\tilde \Gamma$  invariant under (\ref{WGS}),
one shows that $\tilde R=g^{\mu\nu}\,\tilde R_{\mu\nu}$ transforms
like $g^{\mu\nu}$ i.e.  $\tilde R^\prime=\Sigma^{-2}\tilde R$, but $\tilde\nabla_\mu \tilde R$
is not  Weyl covariant: $\tilde\nabla_\mu^\prime \tilde R^\prime\not=\Sigma^{-2}\tilde \nabla_\mu
\tilde R$.

However, there does exist a Weyl gauge covariant formulation \cite{Dirac,DG1}
of this geometry, as required for a  gauge theory. One  defines a
gauge covariant derivative $\hat\nabla_\mu$
of a  tensor field  $T^{\mu_1....\mu_r}_{\nu_1.....\nu_p}$
of space-time charge $\tilde q_T$ with  $T^\prime=\Sigma^{\tilde q_T} T$, then 
\cite{DG1}\footnote{In general  $\tilde q_T=p-r+q_T$ where $q_T$ is
the tangent space charge, see e.g. the review in Section 2 of \cite{CDA2}.}
\smallskip
\bea\label{qq}
\hat \nabla_\mu T
= \big[\tilde\nabla_\mu(\tilde\Gamma) + \tilde q_T\, \w_\mu\big]\, T\qquad
\Ra\qquad \hat\nabla_\mu' T'=\Sigma^{\tilde  q_T}\, \hat\nabla_\mu T.
\eea

\medskip\noindent
where we did not display the indices of the tensor $T^{\mu_1....\mu_r}_{\nu_1.....\nu_p}$.

Since $\hat\nabla_\mu$ depends on the charge $\tilde q_T$ of field $T$,
no connection $\hat\Gamma$  can be associated to $\hat\nabla$
for all fields on  which it acts;
hence, this  Weyl covariant formulation is {\it non-affine}, but it is
{\it metric} since we now have $\hat \nabla_\mu g_{\alpha\beta}=0$.
Thus one can do all calculations directly in this geometry, without going
to a (metric) Riemannian geometry  as done in the past
(for a  modern, rigorous interpretation of Weyl geometry
as a gauge theory see \cite{CDA,AC,CDA2}).

One then defines the Riemann tensor of Weyl geometry $\hat R^\lambda_{\,\,\,\mu\nu\sigma}$,
using $\hat\nabla_\mu$ (instead of $\tilde\nabla_\mu$) in the standard definition
of this tensor: $[\hat\nabla_\mu, \hat\nabla_\nu]\,v^\lambda=\hat
R^\lambda_{\,\,\,\mu\nu\sigma}\,v^\mu$, where $v^\mu=e^\mu_a\,v^a$ is a vector
with vanishing Weyl charge on the tangent space,  $q_{v^a}=0$~\footnote{
This definition can be extended if the tangent-space charge of this vector
is non-zero \cite{CDA2} (section 2).}.
With this, one
can compute the Riemann tensor $\hat R^\mu_{\,\,\,\nu\rho\sigma}$,
Ricci tensor $\hat R_{\mu\sigma}=\hat R^{\lambda}_{\,\,\,\mu\lambda\sigma}$  and Ricci scalar
$\hat R=\hat R_{\mu\sigma} g^{\mu\sigma}$  of Weyl geometry, in terms
of their Riemannian geometry counterparts.
These relations are presented in the Appendix, eqs.(\ref{ss1}),
and will be used later on\footnote{The relation of Weyl covariant (metric) formulation
to the non-metric one is:
$\hat R^\rho_{\,\,\sigma\mu\nu}=\tilde R^\rho_{\,\,\sigma\mu\nu}-\delta_\sigma^\rho\,\hat F_{\mu\nu}$.}.
One also shows \cite{DG1} (eq.A-25) that in this Weyl gauge covariant formulation,
the Weyl tensor $\hat C^\mu_{\,\,\nu\rho\sigma}$ associated to
$\hat R_{\mu\nu\rho\sigma}$ is equal to its Riemannian geometry version
($C^\mu_{\,\,\nu\rho\sigma}$), so  $\hat C^\mu_{\,\,\nu\rho\sigma}=C^\mu_{\,\,\nu\rho\sigma}$.

In the  Weyl gauge covariant (metric) formulation  of Weyl geometry,
under transformation (\ref{WGS}) we have \cite{DG1}
\bea
\label{WGS3}
\quad\hat R^\prime&=&\Sigma^{-2} \hat R, \qquad\qquad\quad
\hat R^{\prime}_{\mu\nu}=\hat R_{\mu\nu}, \qquad\quad\,\,\,\quad
\hat  R^{\prime\,\sigma}_{\,\,\,\mu\nu\rho}=\hat R^\sigma_{\,\,\,\mu\nu\rho},
\nonumber\\[5pt]
\hat\nabla_\mu' \hat R^\prime&=&\Sigma^{-2}\,\hat \nabla_\mu \hat R,\quad\quad
\hat\nabla_\alpha' \hat R_{\mu\nu}'=\hat\nabla_\alpha \hat R_{\mu\nu},\quad\quad
 \hat\nabla_\alpha' \hat R^{\prime\, \sigma}_{\,\,\mu\nu\rho}=
 \hat\nabla_\alpha \hat R^\sigma_{\mu\nu\rho},\,\,\quad
\nonumber\\[5pt]
X^\prime&=&\Sigma^{-4} X, \quad\qquad\quad
X  =\hat R_{\mu\nu\rho\sigma}^2, \, \,\hat R_{\mu\nu}^2,\,\,
 \,\, \hat C_{\mu\nu\rho\sigma}^2,\,\,\, \hat G, \, \,\, \hat F_{\mu\nu}^2,
\eea

\medskip\noindent
Here the square of a tensor denotes contraction by the metric of
indices in the same position.  The field strength of $\omega_\mu$ is
$\hat F_{\mu\nu}\!=\partial_\mu \omega_\nu-\partial_\nu\omega_\mu$,
also invariant under (\ref{WGS}).
$\hat G$ is the Chern-Euler-Gauss-Bonnet term of Weyl geometry
(hereafter Euler term), see Appendix, eq.(\ref{hatG}).

We see now that the curvature tensors/scalar {\it and}
$\hat\nabla_\mu$ acting on them do transform  covariantly under (\ref{WGS}),
with the same Weyl charge as the operator itself. This property
of Weyl geometry operators is similar to the implementation
of (internal) gauge symmetries of the SM with respect to
which  fields and their derivatives transform covariantly.

This formulation of Weyl geometry may be seen as 
 a covariantised version of Riemannian geometry with respect
 to the  gauged dilatation symmetry  \cite{DG1,review};
 since this formulation is metric, one can use it in applications \cite{CDA2},
 compute quantum corrections \cite{DG1}, etc.

Let us add that if one is not familiar with Weyl  geometry, one may just
regard eqs.(\ref{ss1}) as redefinitions of Riemann and Ricci tensors
and scalar of Riemannian geometry, such that these redefined
expressions and their derivative $\hat\nabla_\mu$
transform covariantly, as in eqs.(\ref{WGS3}).

Using the last equation in (\ref{WGS3}),
the action of Weyl gauge theory of gravity (``Weyl quadratic gravity'')
associated to Weyl geometry in
 $d=4$ dimensions, that is invariant under (\ref{WGS}), is  then \cite{Weyl2} (see
 also more recent developments in \cite{Ghilen0,SMW,DG1,CDA2,CDA,AC,review,Lee}) \footnote{
   Action $S_{\bf w}$ is easily extended to $d=4-2\,\epsilon$ dimensions by multiplying
   its  integrand by $\hat R^{d/2-2}$ which does maintain the Weyl gauge symmetry of each term in
   the action.}
\medskip
\bea\label{WWW}
S_{\bf w}=\int d^4 x \,\sqrt{g}\,
\Big\{\frac{1}{4!\,\xi^2}\hat R^2
-\frac{1}{\eta^2} \,\hat C_{\mu\nu\rho\sigma}^2
-\frac{1}{4\,\alpha^2}\, \hat F_{\mu\nu}^2+\hat G
\Big\}
\eea

\medskip\noindent
with perturbative couplings $\xi,\alpha, \eta< 1$; for more on topological terms
like $\hat G$ see \cite{AC,DG1}.

Action (\ref{WWW}) undergoes  a Stueckelberg breaking of Weyl gauge symmetry,
in which $\omega_\mu$ becomes massive and decouples.
One is left at low scales with Riemannian geometry and Einstein-Hilbert action and a
positive cosmological constant \cite{Ghilen0,SMW} (we return to this action later in the text).
Correspondingly, there is  a conserved   Weyl gauge current,
$j_\mu\! \propto\! \hat\nabla_\mu \hat R$ with  $\hat\nabla^\mu j_\mu\!=\! 0$
\cite{Ghilen0,CDA2}, which generalises a similar  current in global scale invariant
theories \cite{F1,F2,F3,F4,Bellido}.

{At geometric level, one can define  a more general
Weyl gauge invariant action  than (\ref{WWW}). Inspired by the
Dirac-Born-Infeld (DBI) action  \cite{BI,D} (for a review \cite{S,T})
one can write a DBI-like action due to Weyl  geometry itself,
in $d$ dimensions. This action is \cite{DBI} } \footnote{
{
The structure of  action (\ref{hh}) and subsequent (\ref{sd}) in the presence of SM,
also has some remote similarity  to string action  (e.g. D3 brane effective action)
\cite{DBI}.
The strings action is related to the integral of the worldsheet area
$\propto\sqrt{\det \gamma_{\alpha\beta}}$ generated by the propagating string.
The counterpart to the induced metric  $\gamma_{\alpha\beta}$ is here 
due not only to  geometry 
(e.g. $\hat R g_{\mu\nu}$ term in (\ref{hh}), with $\hat R$ a
string tension counterpart), but  also due to matter/SM  
that modifies the  geometry/metric (backreaction), see (\ref{sd}).
This excludes related actions like $\int d^d\sigma
{\prod_i\det^{n_i} (A^{(i)}_{\mu\nu})}$,
$\sum_i n_i\!=\!\frac12$, that could  otherwise
give, when expanded, a similar leading order action.}}
\bea\label{hh}
S_{\bf w}'&=&
\int d^d x \,\big\{- \det \big[a_0 \, \hat R\,g_{\mu\nu}+ a_1\, \hat R_{\mu\nu}
  +a_2\,\hat F_{\mu\nu}\big]\big\}^\frac{1}{2}.
\eea

\medskip
    {This action is rather special: unlike all other DBI-like actions, it has
two new features: first, it has a gauge symmetry (Weyl gauge symmetry)
in curved space-time in $d$ dimensions;
each term under $\det$ is invariant, see (\ref{WGS3}), while $a_{0,1,2}$ are some
{\it dimensionless} coefficients. Second, another important feature is that
no UV regulator scale (or field, tension, etc) is
needed here to make this action well-defined in $d=4-2\epsilon$ dimensions.}

A particular expansion of $S_{\bf w}'$ in ratios of (dimensionless) couplings
$a_j/a_0$, $j=1,2$, recovers  in the leading order the  Weyl gauge theory of quadratic
gravity, eq.(\ref{WWW}), while sub-leading orders  might account for some
quantum corrections to (\ref{WWW}), see \cite{DBI}.
The immediate natural question is whether one can extend $S_{\bf w}'$ to
include matter (Standard Model)?

\section{WDBI action:  unification of Gravity and SM}\label{U}

In this section we construct the Weyl-Dirac-Born-Infeld (WDBI) action
of SM and Weyl geometry and study its properties, inspired by action (\ref{hh}).
The  goal is to write a gauge theory  action that  includes  SM  interactions
alongside the gravitational interactions, on equal footing, while respecting both
SM and Weyl gauge symmetries  in  $d=4-2\epsilon$ dimensions.

To achieve this goal, we must identify all  operators constructed from  SM and Weyl
geometry (curvatures) fields, that have  mass dimension 2 (in $d=4-2\epsilon$ dimensions)
and are both SM and Weyl gauge invariant (Weyl charge $q=0$).
 Why  operators of mass dimension 2?
 Using these operators,  the square root of the $d$-dimensional determinant
 of their linear combination (denoted $A_{\mu\nu}$) has mass dimension $d$.
 Therefore,  the associated WDBI action is  automatically dimensionless and
 mathematically well-defined in  $d=4-2\epsilon$ dimensions,
 with dimensionless couplings, {\it without any additional regulator}
 like a DR  scale $\mu$,
 required in all other gauge theories in $d=4-2\epsilon$. 
This  has important implications discussed later.

The WDBI action sets on equal footing SM operators and Weyl geometry operators
($\hat R_{\mu\nu}$, $\hat R$, etc), internal and external gauge symmetries, and gives a 
 unified description, by the gauge principle, of gravity and SM.
 This is a far more general gauge theory action than  the (quadratic) gauge theory
 action of  SM  in Weyl geometry (SMW) \cite{SMW}, see shortly.
 
First let us specify the transformation of SM  scalars $\phi$ and
fermions $\psi$ under (\ref{WGS})
\bea\label{SMfields}
\phi^\prime=\Sigma^{q_\phi}\,\phi,\quad 
\psi^\prime=\Sigma^{q_\psi}\,\psi,\quad
q_\phi=-\frac{1}{2} (d-2), \quad
q_\psi=-\frac{1}{2} (d-1), \quad \Sigma=\Sigma(x)
\eea
%
The Weyl charges of $\phi$, $\psi$ are found from their (invariant) kinetic terms
in curved space-time in $d$ dimensions (see e.g. the appendix in \cite{SMW}).
This is possible since SM with a vanishing Higgs
mass {\it parameter} is scale invariant and gauging this scale symmetry
(to obtain SM with Weyl gauge symmetry) is then immediate \cite{SMW}.
If $d=4$  we have $q_\phi=-1$ and $q_\psi=-3/2$ i.e. Weyl charges
coincide with their inverse mass dimension.
The Weyl gauge covariant derivatives of $\phi$, $\psi$
are  covariantised versions of their Riemannian
version with respect to the gauged dilatation symmetry
and transform covariantly with same charge, as shown below.

\subsection{Weyl invariant operators of mass dimension two}\label{set}

Let us write a set of operators defined by the fields of SM and Weyl conformal geometry,
that in $d=4-2\epsilon$ dimensions have a mass dimension 2  and are
both SM and Weyl gauge invariant
(Weyl charge $q=0$). In doing so, we include operators suppressed by
powers of the Weyl scalar curvature $\hat R$.
Using (\ref{WGS}),  (\ref{WGS3}) and (\ref{SMfields}), the list of such operators includes:

\medskip\noindent
$\bullet$ Weyl geometry operators
\bea\label{inv2}
\hat R\,g_{\mu\nu},\qquad
\hat R_{\mu\nu},\qquad
\hat F_{\mu\nu}.
\eea

\medskip\noindent
$\bullet$ SM gauge sector:
\bea
F^{(1)}_{\mu\nu}, \qquad
F^{(j)}_{\alpha\beta} F^{(j)}_{\rho\sigma} g^{\alpha\rho} g^{\beta\sigma} \hat R^{-1}\, g_{\mu\nu}.
\eea

\medskip\noindent
Here $F^{(1)}_{\mu\nu}$ is the field strength of SM hypercharge field $B_\mu$,
$F^{(1)}_{\mu\nu}=\partial_\mu B_\nu-\partial_\nu B_\mu$,
and $F^{(i)}_{\alpha\beta}\,F^{(i)\,\alpha\beta}$, $i=1,2,3$ are SM
gauge kinetic terms  for U(1),  SU(2), SU(3), in this order.

\medskip\noindent
$\bullet$  Higgs sector: 
\medskip
\bea
\!(\hat\nabla_\alpha H) (\hat\nabla^\alpha H)^\dagger \hat R^{1-d/2}\, g_{\mu\nu},
\,\,\quad
H^\dagger H \,\hat R^{2-d/2}\,g_{\mu\nu},
\,\,\qquad
(H^\dagger H)^2\,\hat R^{3-d}\,g_{\mu\nu},
\eea
where 
\bea\label{HD}
\hat\nabla_\alpha H=\big(
D_\alpha + 
q_H \,\omega_\alpha)\,H,\qquad
q_H=-\frac{1}{2} (d-2).
\eea

\medskip\noindent
Here $D_\alpha H= (\partial_\alpha -i \cA_\alpha) H$
is the SM covariant derivative of the Higgs doublet,
 $\cA_\alpha=(g/2) \vec \sigma.\vec \cA_\alpha +(g^\prime/2)\, B_\alpha$, with
$\vec \cA_\alpha$  the SU(2) gauge boson, $B_\alpha$ the U(1) of hypercharge,
of gauge couplings $g$ and $g^\prime$ respectively.
One  checks that  $\hat\nabla_\alpha H$ transforms covariantly
under  SM and Weyl gauge symmetry with the same charges as $H$.

\bigskip\noindent
 $\bullet$ SM fermionic sector (sum over  SM fermions understood):
\medskip
\bea\label{fff}
 \big(\,
 i\, \overline\psi \gamma^a\,e_a^\alpha\hat\nabla_\alpha\psi\,
 +{\rm h.c.}
 \big)\hat R^{1-d/2} \, g_{\mu\nu},
\eea
with
\bea\label{FD}
\hat\nabla_\alpha \psi=\Big[ D_\alpha +
  q_\psi \,\omega_\alpha +
  \frac12\,\tilde s_\alpha^{ab}\sigma_{ab}\Big]\psi,\qquad q_\psi=-\frac{1}{2} (d-1).
\eea

\medskip\noindent
$D_\alpha$ is the usual SM-covariant derivative of fermions, 
$\sigma_{ab}=(1/4)[\gamma_a,\gamma_b]$,  ($a,b$ are tangent space indices), and
  $\tilde s_\mu^{ab}$ is the spin connection in Weyl geometry; this  has an
expression given by the covariantised version (with respect to gauged dilatations)
of the Riemannian spin connection
$s_\alpha^{ab}=-e^{\lambda\, b}\, (\partial_\alpha e_\lambda^a- \Gamma_{\alpha\lambda}^\nu \,e_\nu^a)$
see e.g. \cite{SMW,CDA}:
\medskip
\bea
\tilde s_\alpha^{ab}=s_\alpha^{ab}\Big\vert_{\partial_\alpha e_\nu^a
  \ra [\,\partial_\alpha+     \,\omega_\alpha]\,e_\nu^a}
=s_\alpha^{ab}+ (e_\alpha^a\,e^{\nu\,b}-e_\alpha^b\,e^{\nu \,a})\,\omega_\nu,
\eea

\medskip\noindent
where we used that $e_\nu^a$ has Weyl charge $q=1$ (half of that of $g_{\mu\nu}$).
Note $\tilde s_\alpha^{ab}$ is invariant under (\ref{WGS}), therefore
$\hat\nabla_\alpha\psi$
transforms covariantly under (\ref{WGS}) with the same Weyl charge as $\psi$.
Further, one notices
$\gamma^\alpha \tilde s_\alpha^{ab} \sigma_{ab} 
= \gamma^\alpha s_\alpha^{ab} \sigma_{ab} +
(d-1) \gamma^\alpha \omega_\alpha$  with $d=4-2\epsilon$;
therefore, in (\ref{FD}) the dependence on $\omega_\alpha$ of the spin connection
is cancelled by that from $q_\psi \, \omega_\alpha$; then
$\gamma^\alpha \hat \nabla_\alpha\psi=\gamma^\alpha \nabla_\alpha\psi$ and
then the expression in (\ref{fff}), invariant under (\ref{WGS}), becomes
\medskip
\bea\label{Rpsi}
 \big(\, i\, \overline\psi \gamma^a\,e_a^\alpha\nabla_\alpha\psi\,  +{\rm h.c.}
 \big)\,\hat R^{1-d/2} \, g_{\mu\nu},
\eea

\medskip\noindent
with  Riemannian  operator  $\nabla_\alpha\!=\! D_\alpha+ (1/2) s_\alpha^{ab}\,\sigma_{ab}$.
So even though they are charged under (\ref{WGS}), in $d=4-2\epsilon$ dimensions
fermions do not couple directly to $\omega_\alpha$ at tree-level  except
through $\hat R$, see eq.(\ref{ss1}) (for $d=4$ see \cite{Kugo,SMW}).

\bigskip\noindent
$\bullet$ Yukawa sector:
\medskip
\bea\label{inv3}
 \big[\,\big(\,\overline \psi_L\,Y_\psi H\,\psi_R +\overline\psi_L Y^\prime_\psi
  \tilde H \psi^\prime_R\,\big) +\textrm{h.c.}\,\big] \,\hat R^{2-3d/4} \,g_{\mu\nu}.
\eea

\medskip\noindent
with $\tilde H=i\sigma_2 H^\dagger$ and $Y$, $Y^\prime$ Yukawa matrices.
It is easily checked that the sum of the Weyl charges of the fields present
is zero and this operator has mass dimension 2.

\bigskip\noindent
    {\bf $\bullet$} Gauge kinetic mixing term ($\omega_\mu$ - hypercharge):
    \medskip
    \bea
     \hat F_{\alpha\beta}  F^{(1)\,\alpha\beta} \hat R^{-1} g_{\mu\nu}.
     \eea

     \medskip\noindent
     This is invariant under SM group; it is also  Weyl gauge invariant, with mass dimension 2.

\bigskip\noindent
    {\bf $\bullet$} Gauge kinetic term of $\omega_\mu$
    \medskip
    \bea\label{19}
     \hat F_{\alpha\beta} \hat F^{\alpha\beta} \hat R^{-1} g_{\mu\nu},
     \eea  
which is also Weyl gauge invariant, with mass dimension 2.

Additional operators of Weyl charge $q=0$ and mass dimension
two are possible and will be discussed later (section~\ref{oc2}).
Note also that we considered operators suppressed at most by one
power of $\hat R$ for $d=4$; higher suppression powers can
be considered,  but they will not introduce new terms in the
leading order action (section~\ref{oc2}).

\subsection{WDBI action in $d=4-2\,\epsilon$ dimensions}\label{wdbi}

Using operators (\ref{inv2}) to (\ref{19}), we  write
a linear combination ($A_{\mu\nu}$) of these and integrate $\sqrt{\det A_{\mu\nu}}$ in
$d=4-2\epsilon$ dimensions. This gives  a version of
Dirac-Born-Infeld action of both SM  and Weyl geometry, which we
call Weyl-Dirac-Born-Infeld action (WDBI). The action is then:
\bea\label{sd}
&& S_{\bf d}=\int d^dx\,\Big[-\det A_{\mu\nu}\Big]^\frac{1}{2},
\\[9pt]
  A_{\mu\nu}&=&
  a_0 \, \hat R\, g_{\mu\nu}+ a_1 \,\hat R_{\mu\nu}+ a_2 \,\hat F_{\mu\nu}
  + a_3\,F_{\mu\nu}^{(1)} + a_4^{(i)} \,F^{(i)}_{\alpha\beta} F^{(i)\,\alpha\beta}\,g_{\mu\nu}\,\hat R^{-1}
  \nonumber\\[4pt]
  &+&
  a_5\,\vert \hat\nabla_\alpha H\vert^2\, \hat R^{1-d/2} \,g_{\mu\nu}
  +a_6 \,\vert H\vert^2 \hat R^{2-d/2} g_{\mu\nu}
  +a_7 \vert H\vert^4\,\hat R^{3-d}\,g_{\mu\nu}
 \nonumber\\[4pt]
  &+& a_8\, \big(
  i\, \overline\psi \gamma^a\,e_a^\alpha\hat\nabla_\alpha\psi+\textrm{h.c.}
  \,\big) \, \hat R^{1-d/2}\, g_{\mu\nu}
 \nonumber\\[4pt]
  &+& a_9\,\big( 
  \overline \psi_L\,Y_\psi\, H\psi_R+\overline \psi_L Y^\prime_\psi\tilde H\,\psi^\prime_R+\textrm{
    h.c.}\big)
    \hat R^{2-3\,d/4}\,g_{\mu\nu},\nonumber
  \nonumber  \\[4pt]
    &+& a_{10}\, \hat F_{\alpha\beta} \hat F^{\alpha\beta} \hat R^{-1} g_{\mu\nu}
    +a_{11}\, \hat F_{\alpha\beta}  F^{(1)\,\alpha\beta} \hat R^{-1} g_{\mu\nu},
    \label{amunu}
    \eea

\medskip
Action $S_{\bf d}$  has both SM and Weyl gauge invariances
in $d=4-2\epsilon$ dimensions, with  dimensionless coefficients $a_0,..., a_{11}$.
Note that no UV regulator, DR subtraction scale $\mu$ or field, etc,  is present
in this action (a scale, if present, would actually break Weyl symmetry). We return
to this issue shortly.
Next, define
\bea
X^\lambda_{\,\,\,\nu}=\frac{g^{\lambda\rho} }{a_0\hat R}\,A_{\rho\nu}- \,\delta_{\,\,\nu}^\lambda,
\eea
\vspace{-0.5cm}

\noindent
and expand $S_{\bf d}$  \footnote{
We use
$\big[\det (1+X)\big]^{1/2}= 1+\frac12 \tr X+
\frac14\,\big[\frac12 \,(\tr X)^2-\tr X^2\big]
+\big[\frac{1}{48} \,(\tr X)^3-\frac18 \,\tr X\,\tr X^2+\frac16\tr X^3\big]
+\cO(X^4)
$}
\bea\label{exp}
S_{\bf d}&=&
\int d^dx\,\sqrt{g}\,
\big(a_0\,\vert \hat R\vert\big)^{d/2}\,
\Big\{ 1
  +\frac{1}{2} \,\tr X
  +\frac{1}{4}\,\Big(\frac12 \,(\tr X)^2-\tr X^2\Big)
  +\cO\Big[\Big( \frac{a_j}{a_0}\Big)^3\Big]\Big\},
  \label{id}
\eea
with $g= -\det g_{\mu\nu}$.
$X^\lambda_{\,\,\,\nu}$  depends on ratios of coefficients, $a_j/a_0$ ($j\!=\!1,..,11$)
assumed to be small  $\vert a_j/a_0\!\vert\ll\! 1$, as required
for phenomenological reasons, that we verify later\footnote{In (\ref{exp}) and
below, to simplify notation  we wrote
$\cO[(a_j/a_0)^3]$ but we actually mean $\cO(a_j a_k a_m/a_0^3)$.}.
We find
\medskip
\bea\label{action-cj}
S_{\bf d}\!\!\!\!&=&\!\!\!\!\!\int d^dx\,\sqrt{g}\,
\,\Big\{
  \hat R^{d/2-2}
  \Big[\, c_0 \,\hat R^2 +c_1 \,\big( \hat C_{\mu\nu\rho\sigma}^2-\hat G\big)
    + c_2\,\hat F_{\mu\nu}^2+c_3\, \hat F^{\mu\nu}\, F_{\mu\nu}^{(1)}
    + c_4^{(j)} \,F^{(j)}_{\mu\nu} F^{(j)\,\mu\nu}
    \Big]
 \nonumber\\[2pt]
  &+& c_5\,\vert\hat\nabla_\mu H\vert^2 + c_6 \,\vert H\vert^2\,\hat R+
  c_7\,\vert H\vert^4\,\hat R^{2-d/2}
  + c_8\,\Big(
  \frac{i}{2} \,\overline \psi_L \gamma^a e_a^\alpha \nabla_\alpha\psi_R
  +\textrm{h.c.}\Big)
  \nonumber\\[2pt]
  &+& c_9\, \big(\overline\psi_L Y_\psi H \psi_R
  + \overline\psi_L\,
 Y^\prime _\psi  \tilde H\,\psi^\prime_R  + \textrm{h.c.}\big) \,\,\hat R^{1-d/4}
+\cO\Big(\frac{1}{\hat R^3}\Big) \Big\}
+a_0^{d/2}\,\cO\Big(\frac{a_i}{a_0}\Big)^3.\label{sd0}
\eea

\medskip\noindent
The dimensionless coefficients $c_j$, $j=1,..,9$ are functions of
$a_k$ ($k=1,...,11$), found in  Appendix, eqs.(\ref{c0}) to (\ref{c9})
and show how  terms in action (\ref{sd}) contribute to (\ref{sd0}).
The terms of coefficients $a_{10}$ and $a_{11}$ are 
redundant in the leading order, since they do not bring new operators in the action.
Similarly for the term of  coefficient $a_7$, but its
presence ensures  coefficient $c_7$ (of $\vert H\vert^4 \hat R^{2-d/2}$)
is independent of $c_6$ (of $\vert H\vert^2 \hat R$), for
phenomenological reasons.

Action (\ref{action-cj}) is brought to canonical form  in Weyl geometry,
shown below, with dimensionless physical perturbative couplings
of gravity $\xi, \eta, \alpha\!<\!1$,  SM couplings
$\alpha_j\!< \!1$, ($j=1,2,3$), non-minimal coupling $\xi_H< 1$, correct
signs of kinetic terms and no gauge kinetic mixing $\omega_\mu$-hypercharge
(investigated elsewhere \cite{SMW});
we assume below $\xi\!\ll\! \eta\!\sim\! \alpha\! <\!1$
for physical reasons detailed later. Then we obtain
\medskip
\bea\label{sdp}
S_{\bf d}\!\!\!\!&=&\!\!\!\!\!\int d^dx\,\sqrt{g}\,
\,\Big\{
  \hat R^{d/2-2}
  \Big[\,\frac{1}{4!\, \xi^2} \,\hat R^2 -\frac{1}{\eta^2}
   \,\big( \hat C_{\mu\nu\rho\sigma}^2-\hat G\big)
   -\frac{1}{4\alpha^2}\,\hat F_{\mu\nu}^2
   -\frac{1}{4\alpha_j^2} \,F^{(j)}_{\mu\nu} F^{(j)\,\mu\nu}
    \Big]
 \nonumber\\[3pt]
  &+& \vert\hat\nabla_\mu H\vert^2 -\frac{\xi_H}{6}\vert H\vert^2\,\hat R
 -\lambda\,\vert H\vert^4\,\hat R^{2-d/2}
+ \Big(
  \frac{i}{2} \,\overline \psi_L \gamma^a e_a^\alpha \nabla_\alpha\psi_R
  +\textrm{h.c.}\Big)
\nonumber\\[3pt]
 & +& \big(\overline\psi_L Y_\psi H\psi_R
+ \overline\psi_L\, Y_\psi^\prime \tilde H\,\psi^\prime_R
+   \textrm{h.c.}\big) \,\,\hat R^{1-d/4}
+\cO\Big(\frac{1}{\hat R^3}\Big)\Big\}
+ a_0^{d/2}\cO\Big(\frac{a_i}{a_0}\Big)^3.
\eea

\medskip\noindent
This is one of the main results of the paper that we discuss in detail shortly (section~\ref{prop}).
First, demanding that coefficients $c_j$  have the values shown in (\ref{sdp}), (\ref{sys}),
we find a solution for coefficients $a_j$ in action (\ref{sd})
that  brings (\ref{action-cj}) to canonical form (\ref{sdp}).
We have that $a_0$, $a_1$  are fixed by the two equations below
\bea\label{a0}
&& \qquad\qquad\qquad\quad
a_0^{d/2}=\frac{1}{\eta^2}\,\frac{16 (d-3)}{d-2}\,\Big(\frac{a_0}{a_1}\Big)^2,
\\[8pt]
&&  \frac{a_0}{a_1}=\frac{-1}{4}( 1\pm\sqrt{1+16 \kappa}\approx \mp\sqrt{\kappa},\quad
  \kappa\equiv\frac{(d-2)}{16(d-1)}\Big[\frac{\eta^2}{24\xi^2}\frac{d-1}{d-3}-1\Big]
   \gg 1.
  \eea

\medskip\noindent
so $a_0\sim \xi^{-4/d}$.
Assuming for simplicity $a_{10}=0$ (this is easily relaxed), then we find $a_2$
\medskip
\bea
    \frac{a_2}{a_1}&=&\!\!\!\frac{d-2}{2}\,\big(-1\pm \sqrt{1-z}\big)\sim \cO(1),
    \quad
    z\equiv\frac{\eta^2}{4\,\alpha^2}\frac{1}{(d-2)(d-3)}.
    \quad    \quad
    \eea

\medskip\noindent
$z<1$ for  $\eta^2<4\alpha^2 (d-2) (d-3)$.  The physical
couplings $\xi$, $\eta$, $\alpha$ in (\ref{sdp}) are then fixed by  $a_{0,1,2}$ above.
The rest of physical couplings are obtained for the following  $a_j$, $j=4,..,11$:
\smallskip
\bea\label{ajs}
\!\!\!\!\!
a_4^{(j)}\!\!\!\!
&=&\!\!\!\!\frac{1}{4\, f}\,\Big[-\frac{a_0^{2-d/2}}{\alpha_j^2}
-a_3^2\,\delta_{j1}\Big],\, (j=1,2,3);\quad
a_5=a_8=a_9=\frac{a_0^{2-d/2}}{f},\quad
a_6=-\frac{\xi_H}{6}\frac{a_0^{2-d/2}}{f}
\nonumber\\[7pt]
\!\!\! a_7\!\!\!\!&=&\!\!\!\!
\frac{-1}{f}\Big[\lambda a_0^{2-d/2}+a_6^2 \frac{d(d-2)}{8}\Big],
\quad\,
a_{11}\!=\!\frac{-a_3}{4 \,f} (2 a_2+a_1 (d-2)),\quad\,
f\equiv \frac{a_0\,d}{2} + a_1\frac{d-2}{4}.\,\label{eqa}
\eea
    
\medskip\noindent
We see that $a_{1,2}\sim a_0 \,\xi\sim \xi^{1-4/d}$ and
$a_j\sim a_0^{1-d/2}\!\sim \xi^{2-4/d}$, ($j=4,..,11$; $d=4-2\epsilon$); next,
we also impose this last relation to  $a_3$, which is possible
since the above $a_{11}$ enforces $c_3=0$, leaving $a_3$  arbitrary.
To conclude,  $\vert a_{1,2}/a_0\vert \sim \xi\!\ll\! 1$,
$\vert a_j/a_0\vert \sim a_0^{-d/2}\!\sim \xi^2\!\ll\! 1$, $j=3,...,11$,
and the convergence of  expansion (\ref{exp})  is then assured for
our solution for $a_k$, giving action (\ref{sdp})  in $d=4-2\epsilon$ dimensions.

\subsection{Properties of  WDBI action}\label{prop}

Action (\ref{sdp}) is still in the Weyl geometry formulation.
To obtain this action in a Riemannian formulation, one  simply
replaces $\hat R$ of Weyl geometry by its Riemannian
expression shown in eq.(\ref{ss1}). All other terms, except $\hat G$,
are unchanged: indeed, $\hat F_{\mu\nu}$ has the same
expression in Riemannian and also in flat case, and
in the Weyl covariant formulation used here
the term $\hat C_{\mu\nu\rho\sigma}^2$ is equal to its  Riemannian version, so
$\hat C_{\mu\nu\rho\sigma}^2=C_{\mu\nu\rho\sigma}^2$, eq.(\ref{ccc}).

Regarding  $\hat G$ (Euler term),
it is a topological term (total derivative) if $d=4$ (hence it does not
affect the equations of motion), but this  changes in $d=4-2\epsilon$ dimensions
(for a discussion see \cite{DG1});
its expression in Riemannian notation is found in (\ref{hatG}) with
$\hat R_{\mu\nu\rho\sigma}$, $\hat R_{\mu\nu}$ and $\hat R$ replaced
by their Riemannian counterparts, eqs.(\ref{ss1}).

Action  (\ref{sdp}) has interesting properties:

\medskip\noindent
{\bf (a)} In  the leading order of $S_{\bf d}$ we obtained Weyl gauge invariant
actions of the SM  and of Weyl  quadratic gravity (eq.(\ref{WWW}))
in $d=4-2\epsilon$ dimensions; there are also  non-minimal couplings
of SM to gravity ($\xi_H$ and those induced by $\hat R$ which contains $\omega_\mu$).
If $d=4$, the geometric part of this action
(first three terms in (\ref{sdp})) recovers the Einstein-Hilbert gravity
after a Stueckelberg mechanism \cite{Ghilen0,SMW}, as
reviewed in the next section.

\medskip\noindent
    {\bf (b)} With $d=4-2\,\epsilon$, we see that in (\ref{sdp})
    the exact WDBI action {\it predicts} that  
the scalar curvature $\hat R^{-\epsilon}$  i.e. geometry acts as the
UV regulator ``scale''\footnote{This requires $\hat R$ be non-zero, see later.} 
for the leading order action of expanded $S_{\bf d}$.
This is possible due to the Weyl gauge covariance of $\hat R$. 
The   leading order action is thus mathematically well-defined and
needs no UV regulator  (field or scale); the regularisation is ``built-in''
exact $S_{\bf d}$. Being Weyl gauge invariant in $d=4-2\epsilon$ dimensions, the leading order
action is Weyl anomaly-free\footnote{In Riemannian
case Weyl anomaly \cite{Duff,Duff2,Duff3,Deser1976,Deser,Englert} appears from
$\mu$-dependent terms with (local) Weyl symmetry broken by regularisation
in $d=4-2\epsilon$ and from  $\mu$-independent Euler term. This situation
changes in  Weyl geometry \cite{DG1} where in $d=4-2\epsilon$,
Weyl gauge symmetry is preserved, with  Euler term  Weyl gauge covariant.},
as discussed in \cite{DG1} with the regularisation derived here.
Actually, at each order in the expansion, $S_{\bf d}$ is Weyl gauge
invariant and Weyl-anomaly free.

\medskip\noindent
{\bf (c)} The  (exact) WDBI  action, being itself Weyl gauge invariant
in $d=4-2\epsilon$ dimensions, is Weyl anomaly-free, too.
Thus,  the WDBI action is a consistent (quantum) gauge theory.
    
If one starts with the WDBI action in $d=4$,
its analytical continuation  to $d=4-2\epsilon$ does not require
a DR  scale $\mu$ - this is replaced by $\hat R$;
the action is then  mathematically
well-defined and Weyl gauge invariant, with no  added UV regulator scale/field.
Quantum calculations can  be performed
in this Weyl gauge invariant phase, respecting all symmetries
of the theory. This shows the power  of Weyl geometry as a gauge theory.

This elegant behaviour is  unique, not seen in  theories
in Riemannian geometry, where a UV regulator  (scale or field)
is necessarily added ``by hand'', to ensure the theory
is mathematically well-defined in $d=4\!-\! 2\epsilon$ dimensions.
In particular, in conformal gravity a dilaton field is added ad-hoc as regulator
to maintain its  symmetry \cite{Englert} in $d=\!4\!-\!2\epsilon$ dimensions.
Finally, unlike here, in string theory local Weyl invariance  (on the Riemannian worldsheet,
not in physical space-time as here) cannot  be preserved at one-loop
by DR scheme which breaks it in $d=2+\epsilon$.
It is restored  by the condition of vanishing Ricci tensor in target space, e.g.
\cite{Tong}. As a side-remark, this condition  may not be necessary if worldsheet
geometry is that of Weyl geometry where this symmetry is natural
in $d$ dimensions, see Appendix.

\subsection{WDBI  action  in d=4 and the broken phase}

Let us consider now the case of $d=4$ dimensions  in action
(\ref{sd}), (\ref{sdp}). We have
\medskip
\bea\label{s4ini}
S_{\bf 4}&=&\int d^4x\,\sqrt{g} \,\Big[-\det A_{\mu\nu}\Big]^{1/2}
\\[5pt]
\textrm{with}\qquad A_{\mu\nu}&=&
  a_0 \, \hat R\, g_{\mu\nu}+ a_1 \,\hat R_{\mu\nu}+ a_2 \,\hat F_{\mu\nu}
  + a_3\,F_{\mu\nu}^{(1)}
  + a_4^{(j)} \,F^{(j)}_{\alpha\beta} F^{(j) \,\alpha\beta}\,g_{\mu\nu} \hat R^{-1}\qquad\qquad
  \nonumber\\[4pt]
  &+&
  a_5\,\vert \hat\nabla_\alpha H\vert^2 \hat R^{-1} \,g_{\mu\nu}
  +a_6 \,\vert H\vert^2  g_{\mu\nu}
  +a_7 \vert H\vert^4\,\hat R^{-1} g_{\mu\nu}
\label{amunup} \nonumber \\[6pt]
  &+&
  a_8\, \big(
  i\, \overline\psi \gamma^a\,e_a^\alpha\hat\nabla_\alpha\psi+\textrm{h.c.}
  \,\big) \,\hat R^{-1}\, g_{\mu\nu}
  \nonumber\\[4pt]
  &+& a_9\,\big( 
  \overline \psi_L\,Y_\psi  H\psi_R+\overline \psi_L Y^\prime_\psi
  \tilde H\,\psi^\prime_R+\textrm{h.c.}\big)
    \,\hat R^{-1}\,g_{\mu\nu}
   \nonumber \\[4pt]
    &+& a_{10}\, \hat F_{\alpha\beta} \hat F^{\alpha\beta} \hat R^{-1} g_{\mu\nu}
    +a_{11}\, \hat F_{\alpha\beta} F^{(1)\,\alpha\beta} \hat R^{-1} g_{\mu\nu}.
\eea
Action (\ref{sdp}) becomes
\medskip
\bea\label{sd2}
S_{\bf 4}\!\!\!\!&=&\!\!\!\!\!\int d^4 x\,\sqrt{g}\,
\,\Big\{
\frac{1}{4!\, \xi^2} \,\hat R^2 -\frac{1}{\eta^2}
   \, \hat C_{\mu\nu\rho\sigma}^2 
   -\frac{1}{4\alpha^2}\,\hat F_{\mu\nu}^2
   \nonumber\\[3pt]
   &-&
   \frac{1}{4\alpha_j^2} \,F^{(j)}_{\mu\nu} F^{(j)\,\mu\nu}
   +\vert\hat\nabla_\mu H\vert^2 -\frac{\xi_H}{6}\vert H\vert^2\,\hat R
 -\lambda\,\vert H\vert^4\,
+ \big(
  \frac{i}{2} \,\overline \psi_L \gamma^a e_a^\alpha \nabla_\alpha\psi_R
  +\textrm{h.c.}\big)\,
\nonumber\\[3pt]
 & +& \big(\overline\psi_L Y_\psi H\psi_R
  + \overline\psi_L\, Y^\prime_\psi
  \tilde H\,\psi^\prime_R  +   \textrm{h.c.}\big) \,\,
+\cO\Big(\frac{1}{\hat R^3}\Big)\Big\}
+ 
\cO\big(\xi\big)
\eea
provided that 
\bea
a_1&=&\pm\frac{2\sqrt{2}}{\eta},\quad
a_0=\frac{-a_1}{4} \big(1\pm\sqrt{1+16\kappa}\big),\quad
\kappa=\frac{1}{24}\Big[ \frac{\eta^2}{8\xi^2}-1\Big]\gg 1,
\quad
\nonumber\\
a_2&=&(-1\pm\sqrt{1-z})\, a_1; \quad z=\frac{\eta^2}{8\alpha^2},
\quad
a_4^{(j)}=\frac{-1}{4\,f} \Big[\frac{1}{\alpha_j^2}-a_3^2\,\delta_{j1}\Big];\quad j=1,2,3.
\nonumber\\
a_5\!\!&=&\!\! a_8=a_9=\frac{1}{f},
  \quad
  a_6=-\frac{\xi_H}{6\,f},
  \quad
  a_7=\Big[-\lambda-\frac{\xi_H^2}{36 \,f^2}\Big]\frac{1}{f},
  \quad
a_{11}=\frac{-a_3}{2 f} (a_2+a_1).
  \eea

\medskip\noindent
where $f=2 a_0+a_1/2$.
The Weyl gauge covariant derivatives of Higgs and fermions in (\ref{sd2})
are immediate from their  expressions in
eqs.(\ref{HD}), (\ref{FD}), (\ref{Rpsi}) evaluated for $d=4$.
The topological term $\hat G$ was removed from $S_{\bf 4}$,
being a total derivative.

As mentioned,  $S_{\bf 4}$ of (\ref{s4ini}) has an immediate
analytical continuation (regularisation),  by replacing
$d=4\ra d=4-2\epsilon$, to obtain the exact WDBI action $S_{\bf d}$ of (\ref{sd})
which is Weyl gauge invariant  and  Weyl anomaly-free.
No regulator is introduced, $\hat R$ plays here this role.

The leading order of $S_{\bf 4}$ contains the SM action
with a mild change in the Higgs sector to make it
Weyl gauge invariant,  with non-minimal gravitational couplings,
plus the  Weyl quadratic gravity action, first line in (\ref{sd2}).
We thus recovered in this leading order the action of SM in Weyl geometry (SMW),
studied in \cite{SMW}.
The exact WDBI action is however  more general and has additional
contributions: these appear in its series expansion as 
sub-leading orders, which are higher dimensional (non-polynomial)
operators, discussed in Section~\ref{slo}.

It is well-known that the leading order action shown in
(\ref{sd2}) has a Stueckelberg  breaking mechanism of  Weyl gauge symmetry
\cite{Ghilen0,SMW}. Since this is relevant for the
sub-leading orders of $S_{\bf 4}$,
we  briefly review this mechanism by considering only
the geometric part  of  $S_{\bf 4}$, shown in (\ref{sd2}) \footnote{
To see the breaking including the effects from SM action shown  in
 (\ref{sd2}), see section 2.5 in \cite{SMW}.}, which is
\medskip
\bea
S_{\bf w}=\int d^4 x\,\sqrt{g}\,
\Big\{\, \frac{1}{4! \,\xi^2} \,\hat R^2 -\frac{1}{\eta^2}
\, \hat C_{\mu\nu\rho\sigma}^2
    -\frac{1}{4\,\alpha^2}\,\hat F_{\mu\nu}^2\Big\}.
    \eea

\medskip
First,   replace in this action $\hat R^2\ra -2 \phi^2 \hat R-\phi^2$,
to obtain a new action which gives an equation of motion for $\phi$ of solution:
$\phi^2=-\hat R$ ($\hat R<0$)\footnote{
$\hat R<0$ is consistent with $R=-12 H_0^2$ ($\Lambda=3 H_0^2$)
obtained for a Friedmann-Robertson-Walker metric.}
which replaced back in the action recovers $S_{\bf w}$;
hence the two actions are equivalent. Next, one goes to the
Riemannian picture, using (\ref{ss1}) for $d=4$, to
write  $\hat R$ in terms of Riemannian scalar curvature $R$.
After some arrangements the action in Riemannian geometry notation
becomes \cite{SMW} 
    \be
    \label{alt2}
      S_{\bf w}\!=\!\!\int\! d^4x \sqrt g\,
\Big\{\frac{-1}{2\xi^2} \Big[ \frac16 \phi^2\,R 
  +(\partial_\mu\phi)^2
  \Big]
-\frac{\phi^4}{4!\,\xi^2} + \frac{\alpha^2 q^2}{8\,\xi^2}\,\phi^2 \Big[\w_\mu
-\partial_\mu \ln\phi\Big]^2\!
-\frac{1}{4}\,F_{\mu\nu}^2\,-\,\frac{1}{\eta^2}\,C_{\mu\nu\rho\sigma}^2
\Big\}
 \ee

 \medskip\noindent
 where
 $F_{\mu\nu}=\partial_\mu\omega_\nu-\partial_\nu\omega_\mu=\hat F_{\mu\nu}$ and we used
 eq.(\ref{ccc}).
The action  remains invariant under  (\ref{WGS}).
By applying transformation (\ref{WGS}) with
$\Sigma=\phi^2/\langle\phi^2\rangle$ one is fixing $\phi$ to its vev,
assumed to exist. Naively, one sets $\phi\ra \langle\phi\rangle$ in $S_{\bf w}$.
In terms of transformed (``primed'')  fields 
the above  action gives in the broken phase
\medskip
\bea
\label{EP}
S_{\bf w}=\int d^4x
\sqrt{g'}  \,\Big[- \frac12\, M_p^2 \, R'
 +\frac12 m_\omega^2 \w_\mu'  \w^{\prime \mu}
 - \Lambda\, M_p^2
 -\frac{1}{4} \, \hat F_{\mu\nu}^{\,\prime \, 2}-\frac{1}{\eta^2}\, C_{\mu\nu\rho\sigma}^2
 \Big],
\eea
where we rescaled $\omega_\mu\ra\alpha\, \omega_\mu$ and
introduced the cosmological constant, Planck scale and the mass of $\omega_\mu$
\bea\label{la}
\Lambda\equiv \frac14\,\langle\phi\rangle^2,\qquad
M_p^2\equiv \frac{\langle\phi^2\rangle}{6\,\xi^2},\qquad
m_\omega^2\equiv 6\, \alpha^2\,M_p^2.
\eea

All mass scales have geometric origin due to the field $\phi$ (from the $\hat R^2$ geometric
term) that generates them \cite{non-metricity,review}. As seen from (\ref{alt2}), 
the gauge field $\omega_\mu$ becomes massive in a Stueckelberg mechanism,
by eating the derivative of $\ln\phi$ field
  which is the would-be-Goldstone of gauged dilatations\footnote{$\ln\phi$
  transforms with a shift under (\ref{WGS}).}.
  This is the Weyl gauge symmetry breaking in the absence of matter.
In the presence of the SM, the new would-be-Goldstone is a mixing (radial direction
in the field space) of $\phi$ and the (neutral) Higgs field ($h$), since now
both contribute to the Planck mass and $m_\omega$ in (\ref{la}); in this sense, in
action (\ref{alt2}) one replaces
$(1/\xi^2) \,\phi^2\ra (1/\xi^2)\,\phi^2+\xi_H h^2.$
The real (neutral) Higgs field is then the angular direction in the field space of
initial $\phi$ and $h$.   For details see \cite{SMW}
(section 2.5 and Appendix C). This ends our review of the  of
breaking of Weyl gauge symmetry.

Since $\Lambda$ and $M_p$ are related,
with $\xi^2\sim\Lambda/M_p^2$, this explains  our initial assumption
$\xi\!\ll\! 1$.
One has $m_\omega\sim M_p$ for $\alpha$ not far below 1, so massive $\omega_\mu$
decouples below $M_p$  and Weyl connection (\ref{tgamma}) and geometry become 
Levi-Civita connection and Riemannian geometry, respectively
\cite{SMW,Ghilen0}\footnote{If one is tuning $\alpha$ to ultra-weak
values ($\ll 1$),  $\omega_\mu$ can in principle
be light (TeV scale or even lower) \cite{SMW}.}.
Further, for $\eta$ near 1, we also have that the spin-two state due to
the $C_{\mu\nu\rho\sigma}^2$ term in the presence of the Einstein term in (\ref{EP}),
has a mass  $\eta\, M_p$ \cite{LAG} and thus it also decouples not far below
$M_p$. Thus, for $\eta\sim  \alpha <1$ not too small,  as we assumed,
one is  left below $M_p$  with the Einstein-Hilbert action,
with  $\Lambda>0$ and SM action with a Higgs sector with a coupling
to $\omega_\mu$. The phenomenology of this action was discussed in \cite{SMW}.

To conclude, the WDBI action in $d=4$, which is Weyl anomaly free,
recovers, in the leading order, a Weyl  gauge invariant action of
SM and Weyl  quadratic gravity.
This gauge symmetry is broken in this order, the massive
Weyl gauge boson and spin-two state decouple near $M_p$ and one then recovers 
Einstein-Hilbert gravity, with $\Lambda\!>\!0$, and the SM action.

\subsection{Sub-leading orders}\label{slo}

What about the sub-leading orders of the expanded WDBI action?
These are Weyl gauge invariant
operators  $\cO(1/\hat R^3)$ (part of $\cO[(a_i/a_0)^2]$)
and $\cO[(a_i/a_0)^3]$,  see (\ref{sd0}), (\ref{sdp}), (\ref{sd2}).

Concerning $\cO(1/\hat R^3)$ terms, their origin is in 
$\tr X^2$ and $(\tr X)^2$; they  arise from multiplying two
SM-like operators of coefficients
$a_j\propto\! a_0^{1-d/2}\!=\xi^{2-4/d}$, ($j=3,4,...,11$; $d=4-2\epsilon$).
They have extra  suppression relative to other terms
$\cO[(a_i/a_0)^2]$ due to mixed contributions SM - gravity,
shown in (\ref{sd0}), (\ref{sdp}), (\ref{sd2}).
Examples of such operators are
\medskip
\bea\label{fff2}
&&\!\!\!\frac{a_4\,a_6}{a_0^2} \vert H\vert^2 F_{\mu\nu}^{(i) \,2} \hat R^{-1-d/2},
\quad
\frac{a_6\,a_7}{a_0^2} \vert H\vert^6 \hat R^{3-3d/2},
\quad
\frac{a_6\,a_9}{a_0}  \vert H\vert^2 \overline\Psi_L Y_\psi H \psi_R \hat R^{2-5d/4},
\eea
%
The coefficients of these operators are  of order $\sim \xi^4$. 
The first operator gives a term in $S_{\bf d}$
\bea\label{rere}
S_{\bf d}\sim
\xi^{2}\int d^d x\sqrt{g}\,\, \frac{\vert H\vert^2 F_{\mu\nu}^{(i) 2}}{\hat R}
\ra
\frac{1}{M_p^2}
\int d^4 x \sqrt{g}\,
 \vert H\vert^2 F_{\mu\nu}^{(i) 2}.
\eea

\medskip\noindent
In the last step we  used the broken phase in $d=4$  with $M_p$ of (\ref{la}).
Relative to the rest of  $\cO(a_j^2/a_0^2)$ operators  that we kept in
the leading order action, this contribution is strongly suppressed by $\xi^2\ll 1$,
(or  by $M_p^2$ in the broken phase). Similar for the other two operators above.
In general, $\cO(1/\hat R^3)$ operators bring $\cO(\xi^2)$ corrections to
the physical couplings of the terms  shown in the leading order action
(recall $\xi^2\sim\Lambda/M_p^2$).

Concerning $\cO[(a_i/a_0)^3]$ operators,  they generate corrections
such as $\cO(a_6^3/a_0^3)$ that contributes to the action a term like
\be
S_{\bf d}\sim
\xi^4 \int d^dx \sqrt{g}\, \,\frac{\vert H\vert^6}{\hat R^{d-3}}
\,\ra\,
\frac{\xi^2}{M_p^2}\int\! d^4x\, \sqrt{g}  \,\, \vert H\vert^6,
\ee
which is more suppressed than (\ref{rere}).
Since such operators respect the gauge symmetry, they may  be
generated as quantum corrections, if one computed these starting from the
leading order action as tree-level action. In other words, the WDBI action
may include some quantum effects, at least on geometric side \cite{DBI}.
To conclude, the  expansion of the WDBI action generates
sub-leading orders which are higher dimensional operators  strongly
suppressed  by powers of gravitational coupling,
$\xi^2\ll 1$ (or by $M_p^2$ in the broken phase).

\subsection{Other corrections}\label{oc2}

The set of Weyl gauge invariant operators of mass dimension 2 in Section~\ref{set},
used to build the WDBI action was {\it minimal}, sufficient to recover in the leading order
a Weyl gauge invariant SM action and Weyl quadratic gravity action.
Additional similar operators  could be present in $A_{\mu\nu}$,
with new dimensionless coefficients. For example another operator is
\medskip
\bea\label{o}
\hat R^{\alpha\beta} \hat F_{\alpha\beta}  \hat R^{-1} g_{\mu\nu} \propto
\hat F_{\alpha\beta} \hat F^{\alpha\beta} \hat R^{-1}\, g_{\mu\nu}
\eea

\medskip\noindent
since the antisymmetric part of  $\hat R_{\alpha\beta}$ is  $\hat F_{\alpha\beta}$.
This operator generates  a gauge kinetic term
for $\omega_\mu$ in the leading order action,
already present in our action;  up to a redefinition of
Weyl gauge coupling, this operator brings no additional physics.
Similarly, the operator obtained from the lhs of (\ref{o}) with $\hat F\!\ra\! F^{(1)}$,
generates a gauge kinetic mixing (hypercharge - $\omega_\mu$),
already discussed  in the leading order and it can also be ignored.

A  more general form of $A_{\mu\nu}$ is 
\medskip
 \bea
 A^\prime_{\mu\nu}&=& A_{\mu\nu}\big[a_k\, g_{\mu\nu}\ra a_k \,(g_{\mu\nu}+z_k\,\kappa_{\mu\nu})\big]
\eea
where $k=4,5,..,11$, and $z_4,...z_{11}$ are new dimensionless coefficients,
with $\kappa_{\mu\nu}\equiv \hat R_{\mu\nu} \hat R^{-1}$
which transforms under (\ref{WGS}) just like the metric.
With the new  $A^\prime_{\mu\nu}$ one shows that the same action
is found in the leading order, up to a redefinition of coefficients
$c_k$, without generating new terms.
One can also extend $\kappa_{\mu\nu}$ to include  corrections to it like
$(1/\hat R^2) \,\hat R_{\alpha\beta} \hat R^{\alpha\beta}  g_{\mu\nu}$,
which  has the same Weyl charge as the metric,
and so on. Such corrections do not bring new terms in the leading order
action discussed, but this may change in higher orders of the expanded action.

\section{Conclusions}\label{C}

In this work we constructed  a general gauge theory beyond
SM and gravity in $d=4-2\epsilon$ dimensions, based on Weyl gauge group
(of dilatations and Poincar\'e symmetries). The natural framework for such gauge
symmetry is Weyl geometry where Weyl gauge symmetry is present  by definition.
We used the   Weyl gauge covariant
(metric!) formulation of this geometry, which we reviewed.
The action we found is a generalised version of the Dirac-Born-Infeld action for
SM and Weyl geometry, which we called Weyl-Dirac-Born-Infeld (WDBI) action.

To find this action, one constructs  a linear combination ($A_{\mu\nu}$) of all
 Weyl-gauge-invariant terms (in $d=4-2\epsilon$) that have
mass dimension two and  are products of SM operators, Weyl
geometry operators and their covariant derivatives.
The space-time integral in $d=4-2\epsilon$  of $\sqrt{\det A_{\mu\nu}}$ 
gives the WDBI action. To our knowledge, this is the most general gauge
theory  of the SM and of gravity based on the Weyl gauge group,
in  $d=4-2\epsilon$  dimensions.

In some sense $A_{\mu\nu}$ can be regarded as a ``new'' space-time metric
and the action as  the new space-time volume.
In $d=4$ dimensions this new action that includes SM, is background independent;
gravity is not a background in which SM exists and this action makes no 
distinction ``brane'' (matter) versus ``bulk'' (gravity) that exists in
string theory. Given the SM contributions to the new metric,
the WDBI action gives a geometric unification of SM and gravity,
removing the distinction matter-geometry or brane-bulk
(with matter fields as ``actors'' on the gravity ``stage''), and
giving instead a {\it relational} approach. There is also a unified picture in
the gauge sense, in that external (gauged dilatations) and internal (SM) gauge
interactions act on equal footing to generate the action.

By construction, the WDBI action is mathematically well-defined in 
$d=4-2\epsilon$ dimensions, with SM and Weyl gauge invariance, and does not
require a UV regulator scale (like a DR scale $\mu$) or field
added ``by hand'', as done in ordinary (quadratic) gauge theories.
Actually, a DR scale $\mu$ would be a problem since it breaks  Weyl gauge symmetry!
The WDBI action actually {\it predicts} that in $d=4-2\epsilon$
the Weyl gauge covariant scalar curvature $\hat R^\epsilon$  i.e. geometry/gravity
acts as a UV regulator for  the $d=4$ theory, as we saw  in particular in a
leading order of its series expansion. This is a special feature of the WDBI action
that maintains Weyl gauge invariance in $d=4-2\epsilon$,
and shows that this action is  more fundamental than ordinary (quadratic)  gauge theories.

This special behaviour is not possible in Riemannian geometry where
Weyl gauge covariance does
not exist; in ordinary  gauge theories a regulator (DR scale $\mu$, etc) is added
by hand. Further, in conformal gravity action a dilaton is also
added by hand as regulator field (to preserve its symmetry in $d=4-2\epsilon$).
Not even in string theory can local Weyl invariance (on Riemannian worldsheet,
not in space-time as here) be respected by regularisation (in $d=2+\epsilon$),
with this symmetry broken by the added DR scale $\mu$;
this symmetry is restored by a condition of vanishing Ricci tensor
but this may not be necessary if the worldsheet geometry is  Weyl geometry
(then  Weyl scalar curvature could act as regulator and preserve the symmetry,
 as here).

Since the WDBI action has manifest Weyl gauge symmetry in $d=4-2\epsilon$ dimensions,
there is no Weyl anomaly, so this action  is a consistent (quantum) gauge theory of gravity.
In the leading order of a series expansion (in $\xi$) of the WDBI
action, one recovers a Weyl gauge invariant version of SM action
plus Weyl (gauge theory of) quadratic gravity;
this theory undergoes a Stueckelberg breaking mechanism in which
the Weyl gauge boson $\omega_\mu$ becomes massive and  Weyl gauge
symmetry is broken. After $\omega_\mu$ decouples below Planck scale,
Riemannian geometry (Levi-Civita connection)
is recovered in the broken phase, together with
the Einstein-Hilbert gravity, SM action and a positive $\Lambda$.

Regarding the  sub-leading orders of the  expansion of WDBI action,
these are operators suppressed by powers of dimensionless gravitational
coupling ($\xi$), with a structure that has some similarities  to
quantum corrections to the leading order action.
In other words, the WDBI action may encode some
quantum corrections. In the broken phase, these operators
are higher dimensional operators suppressed by powers of Planck scale, 
familiar in the SM.

To conclude,  the  WDBI action is a general gauge theory of SM and gravity,
mathematically well-defined and Weyl gauge invariant
in  $d=4-2\epsilon$ dimensions and thus Weyl anomaly-free.
This  is an interesting unified (quantum) description, by
the gauge principle, of SM and gravity, that deserves further study.

\bigskip\bigskip

\begin{center}
  ---------------------------------------------
\end{center}

\newpage

\section*{Appendix}

\def\theequation{A-\arabic{equation}}
\def\thesubsection{A.}
\setcounter{equation}{0}
\def\thefigure{A-\arabic{figure}}
\def\thelabel{A}

\subsection*{$\bullet$ Weyl geometry formulae}

We present some formulae in Weyl geometry and the relation to Riemannian geometry,
in arbitrary $d$ dimensions; in the text, in the WDBI action, we have  $d=4-2\,\epsilon$
($\epsilon\ra 0$).
The relations of  curvature tensors/scalar (with a hat)
in the Weyl gauge covariant formulation of Weyl geometry,  to their Riemannian geometry
counterparts, are found by using their definitions in the text,
see \cite{DG1} (Appendix) and \cite{AC,CDA2}: 
\bea\label{ss1}
\hat R_{\alpha\mu\nu\sigma}
\!\!\!&=&
\!\!\!
R_{\alpha\mu\nu\sigma}
+\Big\{g_{\alpha\sigma} \nabla_\nu\w_\mu- g_{\alpha\nu} \nabla_\sigma \w_\mu
-g_{\mu\sigma}\nabla_\nu \w_\alpha +g_{\mu\nu} \nabla_\sigma\w_\alpha
\Big\}
\nonumber\\
&+&\Big\{ \w^2 (g_{\alpha\sigma} g_{\mu\nu}-g_{\alpha\nu} g_{\mu\sigma} )
+\w_\alpha \,(\w_\nu g_{\sigma\mu}-\w_\sigma g_{\mu\nu})
+\w_\mu (\w_\sigma g_{\alpha\nu} -\w_\nu g_{\alpha\sigma})\Big\}\quad
\nonumber\\
\hat R_{\mu\sigma}\!&=&\!\!\! R_{\mu\sigma} 
\!+ \Big[\frac12 (d-2) F_{\mu\sigma}-(d-2)\nabla_{(\mu} \omega_{\sigma)}
 - g_{\mu\sigma} \nabla_\lambda\omega^\lambda\Big]
\!+(d-2) (\omega_\mu\omega_\sigma
-g_{\mu\sigma} \omega_\lambda\omega^\lambda)
\nonumber\\[7pt]
\hat R &=&g^{\mu\sigma}\hat R_{\mu\sigma}=R-2 (d-1)\, 
\nabla_\mu \omega^\mu -(d-1) (d-2) \,\omega_\mu \omega^\mu.
\eea

\medskip\noindent
with $\hat R_{\alpha\mu\nu\sigma}\!=\! g_{\alpha\lambda} \hat R^{\lambda}_{\,\,\,\,\mu\nu\sigma}$.
Here $R_{\alpha\mu\nu\sigma}=g_{\alpha\lambda} R^\lambda_{\,\,\,\mu\nu\sigma}$,
$R_{\mu\nu}=R^\lambda_{\,\,\,\mu\lambda\nu}$, $R=g^{\mu\nu} \,R_{\mu\nu}$ are the Riemann and
Ricci tensor and scalar  of  Riemannian geometry, respectively, in $d$ dimensions.
The rhs of these equations is in Riemannian notation, with 
 $\nabla_\mu\omega_\nu=\partial_\mu \omega_\nu -\Gamma_{\mu\nu}^\rho\omega_\rho$, and
$\Gamma$ the Levi-Civita connection:
$\Gamma_{\mu\nu}^\rho=(1/2)\,g^{\rho\lambda} (\partial_\mu g_{\nu\lambda}+\partial_\nu g_{\mu\lambda}
-\partial_\lambda g_{\mu\nu}).$

Note  $\hat R_{\mu\nu}-\hat R_{\nu\mu}= 
(d-2) \hat F_{\mu\nu}$, so $\hat R_{\mu\nu}$ is not symmetric if $d\not=2$.
The field strength $ F_{\mu\nu}=\partial_\mu\omega_\nu-\partial_\nu\omega_\mu=\hat F_{\mu\nu}$
 has the same expression as in Weyl geometry.

One shows  that in the Weyl gauge covariant formulation used in this work, 
the Weyl tensor $\hat C^\mu_{\,\,\nu\rho\sigma}$ associated to the Riemann tensor of Weyl geometry
($\hat R^\mu_{\,\,\,\,\nu\rho\sigma}$) is actually equal to its Riemannian counterpart
($C^\mu_{\,\,\nu\rho\sigma}$) \cite{DG1} (eq.A-25)
\bea\label{ccc}
\hat C^\mu_{\,\,\nu\rho\sigma}=C^\mu_{\,\,\nu\rho\sigma}.
\eea

\medskip
In the text we used the following
identities of Weyl conformal geometry (in the ''hat'' notation) 
that are similar to  those of Riemannian geometry, but in a Weyl gauge covariant
form \cite{DG1}, \cite{CDA2}
\medskip
\bea\label{hatG}
\hat G=\hat R_{\mu\nu\rho\sigma}\,\hat R^{\rho\sigma\mu\nu} - 4 \,\hat R_{\mu\nu}\hat R^{\nu\mu}
+\hat R^2
\eea
and
\bea
\hat C_{\mu\nu\rho\sigma}^2=\hat R_{\mu\nu\rho\sigma}\,\hat R^{\rho\sigma\mu\nu}-\frac{4}{d-2}\,
\hat R_{\mu\nu}\,\hat R^{\nu\mu}
+\frac{2}{(d-1)(d-2)}\,\hat R^2,
\eea
giving
\bea\label{hatRmunu}
\hat R_{\mu\nu}\,\hat R^{\nu\mu}=\frac{d-2}{4\,(d-3)} \,(\hat C_{\mu\nu\rho\sigma}^2-\hat G)
+\frac{d}{4\,(d-1)}\,\hat R^2.
\eea

\medskip
The last equation is used in eq.(\ref{id}) to replace  the dependence on the
Ricci tensor ($\hat R_{\mu\nu}$) of Weyl geometry by that on the Weyl tensor of Weyl geometry
in the covariant formulation ($\hat C_{\mu\nu\rho\sigma}$)
since this is identical to the Weyl tensor of Riemannian geometry,
$C_{\mu\nu\rho\sigma}$.

\subsection*{$\bullet$ Coefficients $c_j$}

The coefficients $c_j$ in action (\ref{action-cj}) have the following expressions in terms
of $a_j$ ($d=4-2\epsilon$):
\medskip
\bea\label{c0}
c_0&=& \Big[ a_0^2+ \frac12 \,a_1\,a_0 +a_1^2\,\frac{d-2}{16\,(d-1)}\Big] a_0^{d/2-2}
\\
c_1&=&-\frac{a_1^2\,(d-2)}{16\,(d-3)}\, a_0^{d/2-2}
\\[5pt]
c_2&=&\frac{a_2}{4}\, \Big[\, a_2+a_1\,(d-2)\Big]\, a_0^{d/2-2}+ a_{10}\, f\, a_0^{d/2-2}\label{c2}
\\
c_3&=&  \frac{a_3}{4}\, \Big[ \,2\, a_2 +a_1\, (d-2)\Big]\, a_0^{d/2-2}
+ a_{11}\,f\,a_0^{d/2-2},\label{c3}
\\
c_{4}^{(j)}&=& \Big[ \, a_{4}^{(j)} \,f\, + \delta_{1 j} \,
  \frac{a_3^2}{4}\,\Big]\,a_0^{d/2-2},\quad j=1,2,3,
\\
c_k&=&\Big[ a_k \, f + \frac18\,d\,(d-2)\, \delta_{k 7}\, a_6^2\,\Big]\, a_0^{d/2-2},
\qquad k=5,6,...9. 
\\[7pt]
&&\textrm{with the notation:}\qquad
f= a_0 \frac{d}{2} + a_1 \frac{(d-2)}{4}.
\label{c9}
\eea

\medskip\noindent
The physical couplings in (\ref{sdp}) are related to  $c_j$  
as seen by comparing actions (\ref{sd0}) and (\ref{sdp})
\medskip
\bea\label{sys}
&& c_0=\frac{1}{4!\,\xi^2},\quad c_1=\frac{-1}{\eta^2},\quad c_2=\frac{-1}{4\alpha^2},
\,\,
c_3=0,\quad  c_4^{(i)}=\frac{-1}{4\alpha_i^2},\,\,(i=1,2,3)
\nonumber\\
&& c_5=c_8=c_9=1,\quad c_6=\frac{-\xi_H}{6},\quad
c_7=-\lambda,
\eea

\medskip\noindent
with $\alpha_i$ $(i=1,2,3)$ the gauge  couplings of the SM and
$\alpha$ the Weyl gauge coupling of dilatations.

From (\ref{sys}) with (\ref{c0}) to (\ref{c9})
one finds the values of initial $a_k$ that lead to
physical couplings shown in action (\ref{sdp}); these values  are
presented in eqs.(\ref{a0}) to (\ref{eqa}).

\subsection*{$\bullet$  Weyl invariance in strings}

While this is not important for our study, let us
justify  the last remark at the end of section~\ref{prop}.
Consider the string  action below ($S_s$),
with $\sigma^\alpha$, $g_{\alpha\beta}$ ($\alpha,\beta=1,2$) 
as worldsheet coordinates and metric, respectively.
This  action has {\it local} (rather than gauged)
Weyl invariance i.e. the classical action is invariant
under metric rescalling
$g_{\alpha\beta}\ra g^\prime_{\alpha\beta} =\Sigma^2 g_{\alpha\beta}$;
the difference from gauged Weyl invariance is that,
unlike in (\ref{WGS}), there is no gauge field $\omega_\mu$ in this case ($d=2$).
In a standard notation
\bea\label{string}
S_s&=&\frac{1}{4\pi\alpha^\prime}
\int d^2\sigma
\,
\sqrt{g}\,  \,g^{\alpha\beta} \,
\partial_\alpha X^\mu \,\partial_\beta X^\nu\,G_{\mu\nu}(X).
\label{1}
\eea
At one-loop,  local Weyl symmetry is broken. In a DR scheme in $d=2+\epsilon$,
a regularised $S_s$ is found by
replacing $d^2\sigma\!\ra\! d^{2+\epsilon} \sigma\,\mu^\epsilon$ in (\ref{string}).
The DR scale $\mu$ ensures $S_s$ be dimensionless,
but  initial, classical local Weyl symmetry of  $S_s$  is broken, since
$\sqrt{g}\, g^{\alpha\beta}$ has now a non-zero Weyl charge $d-2=\epsilon$, see (\ref{WGS}).
Then the renormalized  $G_{\mu\nu}(X)$ 
receives a correction of the form
$\alpha^\prime \mathcal R_{\mu\nu}(X)\ln(\Box/\mu^2)$ which is not Weyl invariant,
with $\mathcal R_{\mu\nu}(X)$ the Ricci tensor in target space. Weyl symmetry
is only restored by a (sufficient)
condition of vanishing beta function of $G_{\mu\nu}(X)$, defined
as a derivative with respect to $\ln\mu$, giving
$\alpha^\prime\mathcal R_{\mu\nu}(X)\!=\!0$ \cite{Tong}.

However, if the worldsheet geometry is actually Weyl geometry rather than Riemannian,, 
a Weyl invariant regularised $S_s$ exists,  found by replacing in (\ref{string}):
$d^2\sigma\ra d^{2+\epsilon}\sigma\, \hat R^{\epsilon/2}$
\bea\label{string2}
S_s&=&\frac{1}{4\pi\alpha^\prime}
\int d^{2+\epsilon}\sigma
\,
\sqrt{g}\,  \,g^{\alpha\beta} \,
\partial_\alpha X^\mu \,\partial_\beta X^\nu\,G_{\mu\nu}(X)\,\hat R^{\epsilon/2}
\label{2}
\eea
With $\hat R$ as the worldsheet scalar curvature
of Weyl charge $-2$, (eq.(\ref{WGS3})),
this regularised action, with no regulator scale $\mu$ needed/added,
is now Weyl invariant in $d=2+\epsilon$ and thus,  so are the counterterms and
the renormalised action. One expects at one-loop a
Weyl (gauge) invariant correction 
of the form $\alpha^\prime \mathcal R_{\mu\nu}(X) \ln(\hat\Box/\hat R)$
to $G_{\mu\nu}(X)$. In any case,  there is no need to demand
$\alpha^\prime \mathcal R_{\mu\nu}(X)\!=\!0$ to maintain
local Weyl symmetry in $d$ dimensions, i.e. this is not a necessary condition.
Note from (\ref{ss1}) that for $d=2$:
$\hat R_{\alpha\beta}-\frac12\, \hat R\,g_{\alpha\beta}=R_{\alpha\beta}-\frac12 R g_{\alpha\beta}=0$,
just like in Riemannian case. It may be interesting to study further this
observation.

\end{document}